\documentclass[12pt]{article}   %%LaTeX2e
\input{psfig}
\newcommand{\non}{\nonumber}
\newcommand{\beq}{\begin{equation}}
\newcommand{\eeq}{\end{equation}}
\newcommand{\bea}{\begin{eqnarray}}
\newcommand{\eea}{\end{eqnarray}}
\newcommand{\smallw}{{\scriptscriptstyle W}} %
\newcommand{\smallh}{{\scriptscriptstyle H}} %

\newcommand{\mw}{M_\smallw}
\newcommand{\muw}{\mu_\smallw}
\newcommand{\msq}{\tilde{m}}
\newcommand{\mh}{M_{\smallh^\pm}}

\newcommand{\msd}[1]{m_{\tilde{d}_{#1}}}
\newcommand{\mst}[1]{m_{\tilde{t}_{#1}}}
\newcommand{\msu}[1]{m_{\tilde{u}_{#1}}}
\newcommand{\mug}{\mu_{\tilde g}}
\newcommand{\msbar}{\overline{\rm MS}}

\def\al{\alpha_s}
\def\br{{\rm BR}(B\to X_s \gamma )}

\def \ap   {a_+}
\def \am   {a_-}
\def \ct   {\cos \theta_{\tilde{t}}}
\def \st   {\sin \theta_{\tilde{t}}}
\def \cdt   {\cos^2 \theta_{\tilde{t}}}
\def \sdt   {\sin^2 \theta_{\tilde{t}}}
\def \Yd   {\frac{m_d}{\sqrt{2} \mw \cos \beta}}
\def \Yt   {\frac{m_t}{\sqrt{2} \mw \sin \beta}}
\def \mt   {m_t}

\def \mb   {m_b}
\def \mg   {m_{\tilde{g}}}

\def \MS   {\overline{\mbox{MS}}}

\def\slash#1{\setbox0=\hbox{$#1$}#1\hskip-\wd0\dimen0=5pt\advance
       \dimen0 by-\ht0\advance\dimen0 by\dp0\lower0.5\dimen0\hbox
         to\wd0{\hss\sl/\/\hss}}
\def\gequiv{\raise 0.4ex \hbox{$>$} \kern -0.7 em \lower 0.62 ex \hbox{$\sim$}}
\def\gappeq{\mathrel{\rlap {\raise.5ex\hbox{$>$}}
{\lower.5ex\hbox{$\sim$}}}}
\def\ijmp#1#2#3{{\it Int. Jour. Mod. Phys. }{\bf #1 }(19#2)~#3}
\def\pl#1#2#3{{\it Phys. Lett. }{\bf B#1~}(19#2)~#3}

\def\prl#1#2#3{{\it Phys. Rev. Lett. }{\bf #1~}(19#2)~#3}

\def\pr#1#2#3{{\it Phys. Rev. }{\bf D#1~}(19#2)~#3}
\def\np#1#2#3{{\it Nucl. Phys. }{\bf B#1~}(19#2)~#3}

\newenvironment{appendletter}
 {
  \typeout{ Starting Appendix \thesection }
  \setcounter{equation}{0}
  
 }{
  \typeout{Appendix done}
 }

\begin{document}              

%%%%%%%%%%%%%%%%%%%%%%%%%%%%%%% titlepage %%%%%%%%%%%%%%%%%%%%%%%%%%%%%%%%%%%%
\begin{titlepage}
\begin{flushright}
        \small
        CERN-TH/98-177\\
        DFPD-98/TH/25\\
        TUM-HEP-313/98\\
        hep-ph/9806308\\
        June 1998
\end{flushright}

\begin{center}
\vspace{1cm}
{\large\bf Next-to-Leading QCD Corrections to $B\to X_s \gamma$\\
in Supersymmetry}

\vspace{0.5cm}
\renewcommand{\thefootnote}{\fnsymbol{footnote}}
{\bf    M.~Ciuchini$^a$,
        G.~Degrassi$^b$,
        P.~Gambino$^c$,
        G.F.~Giudice$^d$\footnote
                {On leave of absence from INFN, Sez. di Padova, Italy.}}
\setcounter{footnote}{0}
\vspace{.8cm}

{\it
   $^a$ Dipartimento di Fisica, Universit{\`a} di Roma Tre and INFN,\\
   Sezione di Roma III, Via della Vasca Navale, 84, I-00146 Roma,
   Italy\\
\vspace{2mm}
        $^b$ Dipartimento di Fisica, Universit{\`a}
                  di Padova, Sezione INFN di Padova,\\ 
                  Via F.~Marzolo 8 , I-35131 Padua, Italy\\
\vspace{2mm}
        $^c$ Technische Universit{\"a}t M{\"u}nchen,
                Physik Dept., D-85748 Garching, Germany\\
\vspace{2mm}
        $^d$ Theory Division, CERN CH-1211 Geneva 23, Switzerland \\
\vspace{1cm} }

{\large\bf Abstract}

\vspace{.5cm} 
\end{center}
We compute the QCD next-to-leading order matching conditions of the
(chromo)-magnetic operators relevant for $B\to X_s \gamma$ in supersymmetric
models with minimal flavour violation. The calculation is performed under the
assumption that  the charginos and one stop are lighter than all  other squarks
and the gluino. In the parameter region where a light charged Higgs boson is
consistent with measurements of BR$(B\to X_s \gamma)$, we find  sizeable
corrections to the Wilson coefficients. As a consequence, there is a
significant reduction of the stop-chargino mass region where the supersymmetric
contribution has a large destructive  interference with the charged-Higgs boson
contribution. 
\noindent

% PACS numbers:

\end{titlepage}
\newpage
\eject
%%%%%%%%%%%%%%%%%%%%%%%%%%%%%%%%%%%%%%%%%%%%%%%%%%%%%%%%%%%%%%%%%%%%%%%%%%%%%%%

\section{Introduction}

The inclusive decay rate for $B\to X_s \gamma$ has been first measured
by CLEO with the result $\br =(2.32 \pm 0.57_{\rm stat} \pm 0.35_{\rm syst}
)\times 10^{-4}$~\cite{cleo}. Recently, a preliminary new result based on
about 30\% more data has been presented by the collaboration, 
$\br =(2.50 \pm 0.47_{\rm stat} 
\pm 0.39_{\rm syst}
)\times 10^{-4}$~\cite{cleo2}. The same process has also been measured by
ALEPH at LEP, with the result $\br =(3.11 \pm 0.80_{\rm stat} \pm 
0.72_{\rm syst}
)\times 10^{-4}$~\cite{aleph}.

There has been  significant theoretical effort in refining the prediction
of $\br$ in the Standard Model (SM). Calculations are now available for
the next-to-leading order (NLO) corrections to the anomalous 
dimensions~\cite{misiak}, the matrix elements~\cite{ali}, and the matching
conditions of the Wilson coefficients~\cite{wils,noi}, for the leading
non-perturbative effects~\cite{nonpert}, and for the QED 
corrections~\cite{marc,kagan}. In particular the QCD NLO corrections are
important since they reduce the large scale dependence of the leading
result~\cite{unc}, which amount 
to a 30\% uncertainty in the leading order (LO) theoretical
prediction. Recently, Kagan and Neubert~\cite{kagan} 
have argued that, at the NLO, the 
contributions to $\br$ from the different Wilson coefficients exhibit
a larger scale dependence than the total result, signaling that the
theoretical uncertainty may have been underestimated in previous 
literature~\cite{misiak,buras2,noi}. 
Nevertheless, even considering this effect, they find that the theoretical
error can be evaluated to be about 10{\%}. Combining the different theoretical
studies, the recent complete analysis in ref.~\cite{kagan} gives the SM
prediction
$\br = (3.29\pm 0.33)\times 10^{-4} \times {\rm BR}(B\to X_c e \bar \nu )
/10.5\% $.

While the theoretical prediction is made for the total rate of $\br$,
the experimental data refer only to events with photon energies between
2.2 and 2.7 GeV. The extrapolation from the data to the total rate 
introduces further theoretical uncertainties from the calculation of the
photon energy spectrum~\cite{spec}. This aspect has been recently emphasized by
Kagan and Neubert~\cite{kagan} who have recomputed the spectrum and 
extrapolated from the CLEO high-energy photon data the total rate $\br =
(2.66\pm 0.56_{\rm exp} {}^{+0.43}_{-0.48~{\rm 
th}})\times 10^{-4}$. This is compatible
with the SM within one standard deviation.

To summarize, although the SM prediction is still higher than the
CLEO result, the discrepancy seems no longer statistically 
significant. We should however wait for the forthcoming CLEO
analysis and future studies at LEP to further clarify the situation.
Meanwhile, we use the conservative CLEO upper limit of
$\br < 4.2\times 10^{-4}$ at 95\% C.L. in our numerical analysis.

The inclusive decay $B\to X_s \gamma$ is a particularly interesting probe
of physics beyond the SM, since it is determined by a flavour-violating
loop diagram. Because of this welcome sensitivity on new physics, it is
important to determine the predicted rate with sufficient accuracy in a
variety of models. In the case of two-Higgs doublet models,
results for the next-to-leading order (NLO) 
QCD corrections have been presented in refs.~\cite{noi,strum,borz}.
In these models, a new contribution arises from charged-Higgs exchange which
always increases the SM prediction for $\br$. As a consequence, 
strong limits
on the charged-Higgs mass can be derived. These limits are quite dependent
on the treatment of theoretical errors and on the amount of discrepancy
between theoretical and experimental results~\cite{noi,borz}.

Here we want to extend our
previous analysis of QCD NLO effects in $\br$ to the case of supersymmetry. 
Refining the theoretical calculation in supersymmetry~\cite{bert,bar} (for
reviews see ref.~\cite{bsg})
is important
because $\br$ provides a very stringent constraint on the parameter space
of the model. Moreover in supersymmetry it is possible to evade the strong
limits on the charged Higgs mass, since the chargino contribution can
interfere destructively and reduce $\br$~\cite{bar}. Our result allows us
to give a more reliable estimate of the model parameters necessary to
achieve this cancellation.

We have not tried to perform a complete analysis in the most general
supersymmetric model,
an effort, we feel, that  is 
penalized by the fact that the final result will be very involved and 
will depend upon many unknown 
parameters. Instead, we try to focus on what we believe is the most interesting
part of the parameter space for $B \to X_s \gamma$. Specifically, we
concentrate on the case in which the flavour violation is completely
dictated by the
Cabibbo-Kobayashi-Maskawa (CKM)
angles (insuring predictability) and in which the charginos
and a scalar
partner of the top are light (insuring a sizeable new
contribution).

The paper is organized as
follows. In sect.~2 we define our assumptions on the supersymmetric parameters
and discuss the underlying hypotheses.
Section~3 contains the analytical results for the 
supersymmetric corrections to the relevant Wilson coefficients at the
matching scale. A numerical analysis of $\br$ in supersymmetry at the
NLO is presented in sect.~4.

\section{Minimal Flavour Violation and Heavy Squark-Gluino Effective Theory}

In this section we describe our main theoretical assumptions on the
supersymmetric model: {\it i)} the minimal flavour violation and
{\it ii)} the heavy squark-gluino effective theory that describes 
chargino, stop, and the SM degrees of freedom with other supersymmetric
particles integrated out.

In the supersymmetric extension of the SM with general soft-breaking terms,
there are a variety of new sources of flavour violation. The 
supersymmetry-breaking squark masses and trilinear terms lead to a
mismatch in flavour space between quark and squark mass eigenstates. 
The flavour violation is then described by a large number of mixing
angles, which cannot be determined theoretically. All we know is that
present measurements and limits on various flavour-violating processes
provide quite stringent bounds on these mixing angles (see {\it e.g.}
ref.~\cite{mas}).

A more predictive case is what we will call {\it minimal flavour violation}, 
in which all flavour transitions occur only in the charged-current sector
and are determined by the known CKM mixing angles. This is indeed
the case in several theoretical schemes in which the communication of the
original supersymmetry breaking to the observable particles occurs via
flavour-independent interactions. 
In many of these schemes (especially when supersymmetry breaking occurs at 
low-energy scales) the departure from the minimal flavour violation 
hypothesis, caused by quantum effects, is rather small. Our assumption
can then be justified in gauge-mediated models~\cite{gau} (for a review,
see ref.~\cite{rev}) and in certain
classes of supergravity theories. 
At any rate, it plausibly corresponds to 
the unavoidable flavour violation, present in any supersymmetric model.

Within this class of models, we focus on the special case of interest 
for $B\to X_s \gamma$, in which one stop is considerably lighter than
the other squarks. In order to preserve the successful fit of the
electroweak precision measurements, we assume that the light stop is
predominantly right-handed~\cite{fit}. This assumption can also be 
theoretically justified
by the observation that the renormalization-group evolution of the
squark mass parameters indeed pushes the right-handed stop mass to
smaller values. Therefore, we concentrate on the case in which the
supersymmetry-breaking parameters in the up-squark sector are flavour
diagonal (but not necessarily universal) in the basis in which the
corresponding up-type quark mass matrix is diagonal. The flavour-violating
stop interactions arise only from charged-current effects and are
completely determined by the CKM angles. The stop mass
matrix is given by
\beq
m^2_{\tilde t}=\pmatrix{
{ m^2_{\tilde t_L}}+m^2_t+
(\frac{1}{2}-\frac{2}{3}\sin^2
\theta_W)\cos 2\beta M_Z^2
          & m_t( A_t-\mu\cot\beta)\cr
m_t( A_t-\mu\cot\beta)&
  m^2_{\tilde t_R}+m^2_t+\frac{2}{3}
\sin^2\theta_W\cos 2\beta M_Z^2  }\, .
\label{tmass}
\eeq
Here $m^2_{\tilde t_R}$, $m^2_{\tilde t_L}$, and $A_t$ are 
supersymmetry-breaking parameters, $\mu$ is the Higgs mixing mass and
$\tan \beta$ is the ratio between the two Higgs vacuum expectation values. 
We also assume that the
charginos are lighter than gluinos and heavy squarks. These
approximations considerably simplify the calculation and the final
expressions, and they are appropriate to identify the leading supersymmetric
contribution. 

To implement this scenario we assume the following mass hierarchy
\beq
\mug\sim {\cal {O}}( \mg , m_{\tilde q}, \mst{1}) \gg
\muw \sim {\cal {O}}( \mw  ,\mt , m_{\chi^\pm}, \mst{2}) \gg \mb
\gg \Lambda_{QCD}.
\label{eq:scales}
\eeq
Here and in the following $\muw$ is the mass scale of the 
 charginos ($\chi^\pm$),
and of the lighter stop ($\tilde t_2$), to
be identified with the ordinary electroweak scale. The charged Higgs
($H^\pm$) mass is not constrained.
The scale $\mug$, 
characteristic of all other strongly-interacting supersymmetric
particles (squarks and gluinos), 
is assumed to be larger, say of the order of the TeV. We
compute the QCD NLO corrections to the Wilson coefficients keeping only
the first order in an expansion in $\muw /\mug$.

As usual, the presence of different mass scales
allows us to use a stack of effective theories, obtained by integrating out
of the theory, at each matching scale, the heavy degrees of freedom.
The effective theory just below the scale $\mug$
is particularly simple. 
The only
dimension-five operators involving supersymmetric particles 
 are the chargino (chromo)-magnetic dipole
moment and the operator
$\bar t~t~ \tilde t_2^\star ~ \tilde t_2$,
obtained by integrating out the gluino.
However, an explicit calculation shows that these operators do not
contribute to the NLO matching conditions of magnetic-dipole operators
at the scale $\muw$. Therefore, to our end, the inclusion of the
leading $\muw/\mug$
corrections does not require any new operator other than renormalizable
ones.

The next step is the running of the intermediate effective Hamiltonian
between $\mug$ and $\muw$.
Although the relevant anomalous dimensions are known~\cite{anlauf},
log resummation is not really needed, given the smallness of the relevant
log terms, $\alpha_s(\muw)/4\pi\log(\mug^2/\muw^2)\sim 0.01$--$0.05$ for
typical values of the supersymmetric masses. 

Non-resummed NLO effects of heavy supersymmetric particles,
including $1/\mug$ corrections, are simply given by the {\it one-loop}
Feynman diagrams, containing these particles in the loop, which contribute
to renormalizable operators in the intermediate
effective theory. In other words, we need to compute only corrections
involving gluinos and heavy squarks to the coupling
constants appearing in the  vertices involving the  chargino, 
$\chi_a^- \bar b \tilde t_2$, the $W$ vector boson, $W^- \bar q q^\prime$, 
the charged physical, $H^- \bar qq^\prime$, and unphysical, 
$\phi^- \bar qq^\prime$, scalars.
Heavy ${\cal O}(\mug)$ particle effects renormalize the masses of  
${\cal O}(\muw)$ particles, but these
effects are reabsorbed through the definition of renormalized 
masses.  We use on-shell masses for squarks. Quark masses
are also defined on-shell, except corrections not involving supersymmetric
particles, which are subtracted in the $\msbar$ scheme. This definition
simplifies the insertion of supersymmetric contributions into existing
RG-evolution formulae, where $\msbar$ running quark masses are commonly
used.
Because of our assumption that the light stop is mainly right-handed,
the stop mixing angle  $\theta_{\tilde t}$ is small. For instance,
taking the supersymmetry-breaking
parameter $A_t$ in eq.~(\ref{tmass}) of the order of $\mug$, yields
$\theta_{\tilde t} \sim {\cal O} (\muw / \mug )$. 
In this framework, $1/\mug$ corrections multiplied by $\st$ factors should
be regarded as of higher order.

Results for the renormalized vertices in the
intermediate effective theory are collected in appendix A. Notice
that the $\chi_a^- \bar b \tilde t_2$ vertex renormalization
is infinite. This is not surprising, since supersymmetry and 
the GIM cancellation are
spoiled in the intermediate effective theory, where the heavy squarks and
gluinos have
been integrated out. This divergence cancels out in the matching of the
magnetic operators at the scale $\muw$ against a corresponding divergence
generated by the insertion of the chargino vertex into two-loop
diagrams. This cancellation actually provides a check for the results of the
two-loop calculation. However, the presence of the divergence calls
for a regularization. We choose the na\"\i ve dimensional regularization
(NDR), to be consistent with the calculation of the anomalous
dimension matrix~\cite{misiak}. However, it is known that NDR breaks
supersymmetry. In particular, the renormalizations in NDR
of the gauge boson and gaugino interactions with matter, as well
as the Higgs boson and higgsino interactions, are different and manifestly
violate supersymmetry. Supersymmetric Ward identities are restored
with appropriate shifts of the gauge and Yukawa couplings in the
the $\chi_a^- \bar b \tilde t_2$ vertex~\cite{susyreg}, 
denoted as $\eta_Y$ and $\eta_g$ in 
appendix A. These shifts correspond to the difference of using NDR versus
dimensional reduction (DR)~\cite{sie}, a regularization that preserves
supersymmetry.

The formulae in  appendix A are given in terms of few functions of the ratios
$m_{\tilde q}^2/\mg^2$. Notice however that, whenever $\mst{2}^2/\mg^2$
appears, only terms up to ${\cal O}(\mst{2}/\mg)$ should be retained in the
corresponding functions to be consistent with the operator product
expansion.
Finally, we also keep the leading contribution in the sbottom mixing angle,
since the left-right mixing term $(A_b -\mu \tan \beta )\,m_b$ becomes
important for large values of $\tan \beta$.

\section{Wilson Coefficients}
This section contains the result for the NLO supersymmetric contributions to 
the Wilson coefficients $C_{i}^{eff}(\mu_W)$. Following the discussion
of the previous section, the calculation of the NLO corrections to
 $C_{7}^{eff}(\mu_W)$ and $C_{8}^{eff}(\mu_W)$ can be divided in two parts:
i) the contribution of the heavy particles at the scale $\mug$ that will
appear as renormalization of the coupling constants in the LO diagrams. 
ii) The contribution of the intermediate-scale particles that requires the 
computation of the two-loop gluonic corrections to the LO supersymmetric 
diagrams involving the charginos and the light stop. Concerning the
latter two strategies are at hand. One can match matrix elements of
operators belonging to a basis obtained enforcing the equations of motion,
a procedure that however requires an asymptotic expansion of the relevant
diagrams in the external momenta. 
Alternatively, one can use a larger off-shell operator basis and 
perform the matching on off-shell matrix elements. In this case, one can use
the freedom  of the  off-shell status to  choose a suitable kinematical 
configuration such that the various 
Feynman diagrams can be evaluated using ordinary Taylor expansions in the
external momenta. This second strategy,  already successfully
applied by us in ref.~\cite{noi} to the calculation of the QCD corrections
to the matching conditions of the $\Delta
B =1$ magnetic and chromo-magnetic operators in the SM and
in two-Higgs doublet models, has also been employed in this calculation.
Concerning the technical details we refer to ref.~\cite{noi}.

In the supersymmetric model under consideration
we can organize the Wilson coefficients
of the operators entering the effective Hamiltonian in the following way: 
\bea
C^{eff}_i(\muw)& = &C^{(0)eff}_i(\muw) + \delta^H C_{i}^{(0)eff}(\muw)
 + \delta^S C_{i}^{(0)eff}(\muw)  \non\\ &+& 
\frac{\al(\muw)}{4 \pi} \left[ C^{(1)eff}_i(\muw) + 
\delta^H C_{i}^{(1)eff}(\muw)
 + \delta^S C_{i}^{(1)eff}(\muw) \right] 
\label{wc1}
\eea
where $C^{(k)eff}_i(\muw)$ represents the SM contribution ($k=0,1$),
$ \delta^H C_{i}^{(k)eff}(\muw)$ the additional terms present 
in a two-Higgs doublet model, 
while $\delta^S C_{i}^{(k)eff}(\muw)$ contains the
contribution from supersymmetric particles. Explicit
expressions for $C^{(k)eff}_i(\muw)$ and $ \delta^H C_{i}^{(k)eff}(\muw)$
can be found in sections 4 and 5 of ref.~\cite{noi}\footnote{In ref.~\cite{noi} 
$\delta^H C_{i}^{(k)eff}$ are indicated as $ \delta \,C_{i}^{(k)eff}$.}. 
Concerning the remaining contributions,  the LO $\delta^S C_i^{(0)eff}(\muw)$
terms are given by~\cite{bert}
\bea
\delta^S C_i^{(0)eff}(\muw)&=& 0 \ \ \ i=1,...,6\non\\
\delta^S C_{7,8}^{(0)eff}(\muw)&=&
 \sum_{j=1,2}\left[ 
\frac2{3} \frac{\mw^2}{\msq^2} 
\tilde{V}_{j1}^2  
F^{(1)}_{7,8}(z_j) +
\frac{\tilde{U}_{j2}}{\sqrt{2}\cos\beta} \frac{\mw}{m_{\chi_j}} 
\tilde{V}_{j1} F^{(3)}_{7,8}(z_j) \right. \non\\ &&
- \frac2{3}  t_{1j}^2 
\frac{\mw^2}{\mst{1}^2} F_{7,8}^{(1)}(y_{1j}) -
\frac{\tilde{U}_{j2}}{\sqrt{2}\cos\beta} \frac{\mw}{m_{\chi_j}} 
t_{1j} \ct F^{(3)}_{7,8} (y_{1j}) \non\\ &&
 \left. - \frac2{3}  t_{2j}^2  
\frac{\mw^2}{\mst{2}^2} F_{7,8}^{(1)}(y_{2j})
- \frac{\tilde{U}_{j2}}{\sqrt{2}\cos\beta} \frac{\mw}{m_{\chi_j}} 
 t_{2j} \st F^{(3)}_{7,8} (y_{2j}) \right]
\label{wc2}
\eea
where the functions $F^{(1)}_{7,8}$ are defined in  eqs.~(29)--(30) of
ref.~\cite{noi} and
\bea
F^{(3)}_7(x)&=& 
  {\frac{5 - 7\,x}{6\,{{\left(  x-1 \right) }^2}}} + 
   {\frac{x\,\left( 3\,x-2 \right) }
     {3\,{{\left(  x -1\right) }^3}}}\ln x\non\\
F^{(3)}_8(x)&=&
  {\frac{1 + x}{2\,{{\left( x-1 \right) }^2}}} - 
   {\frac{x}{{{\left( x-1 \right) }^3}}}\ln x .
\label{wc5}
\eea
We have defined
\beq
t_{1j}= \tilde{V}_{j1} \ct
-\tan \theta_{\tilde{t}} Y_j;~~~~~~~
t_{2j}= \tilde{V}_{j1} \st + Y_j;
\label{wc4}
\eeq
\beq
z_j= \frac{\msq^2}{m_{\chi_j}^2};~~~~~ 
y_{kj}= \frac{\mst{k}^2}{m_{\chi_j}^2};~~~~~~~~
Y_j = \frac{\tilde{V}_{j2} \ct}{\sqrt{2}\sin\beta} 
\frac{\bar{\mt}(\mu_\smallw)}{\mw}~~,
\label{wc3}
\eeq
where $\bar{\mt}(\mu_\smallw)$ is the top-quark running mass at the scale 
$\mu_\smallw$.

The stop eigenstates ${\tilde t}_1=\cos \theta_{\tilde t} 
\,{\tilde t}_L
+\sin \theta_{\tilde t} \,{\tilde t}_R$ and 
${\tilde t}_2=-\sin \theta_{\tilde t} \, {\tilde t}_L
+\cos \theta_{\tilde t} \,{\tilde t}_R$ 
have mass eigenvalues $m_{{\tilde t}_1}$ and
$m_{{\tilde t}_2}$ and we have taken all other squarks to be degenerate
with mass $\tilde m$. The two matrices $\tilde{U}$ and $\tilde{V}$ 
diagonalize the chargino mass matrix  according to 
($M$ is the weak gaugino mass)
\beq
\tilde{U} \pmatrix{ M & \mw \sqrt{2} \sin \beta \cr
              \mw \sqrt{2} \cos \beta & \mu } \tilde{V}^{-1}\, 
\eeq
and are assumed to be real. Notice that
 in eq.~(\ref{wc2}) and henceforth  the scalar quark masses are understood
as on-shell.

The last ingredients needed for a complete NLO calculation in our
supersymmetric
model are $\delta^S C_i^{(1)eff}(\muw)$. For the current-current and
penguins operators we have
\bea 
\delta^S C_i^{(1)eff}(\muw)&=&0 \ \ \ \ i=1,2,3,5,6 
\label{wc6} \\
\delta^S C_4^{(1)eff}(\muw)&=&  \frac{\mw^2}{m_{\tilde{t}_2}^2}
\sum_{j=1,2}  t_{2j}^2   
 E_\chi(y_{2j})  \label{wc7}  
\eea
with~\cite{GG}
\beq
E_\chi(x)= \frac{x\,(11-7 x+ 2 x^2)}{18(x-1)^3} - \frac{x}{3(x-1)^4}\ln x .
\label{wc11}
\eeq
The $O(\alpha_s)$ corrections to the coefficient 
of the (chromo-)magnetic operator can be
divided into 4 pieces 
\bea 
\delta^S C_{7,8}^{(1)eff}(\muw)&=& \delta^\chi C_{7,8}^{(1)eff}(\muw) +
\delta^\smallw C_{7,8}^{(1)eff}(\muw) + 
\delta^{\phi_1} C_{7,8}^{(1)eff}(\muw) \non\\
& +& \delta^{\phi_2} C_{7,8}^{(1)eff}(\muw) ~~.
\label{wc8}
\eea
Here $\delta^\chi C_{7,8}^{(1)eff}$ represents the
chargino contribution while 
$(\delta^W,\, \delta^{\phi_1}, \delta^{\phi_2})
 C_{7,8}^{(1)eff}$ take into account the renormalization effects due
to the O($\mug$) heavy particles  
in the $W$, physical and unphysical charged scalar couplings, respectively. 
We find
\bea
\delta^\chi C_{7}^{(1)eff}(\muw)&=&
\sum_{j=1,2} 
\left[  
 t_{2j}^2  \frac{\mw^2}{\mst{2}^2}
 \left( G^{\chi,1}_{7}(y_{2j}) + 
\Delta^{\chi,1}_7(y_{2j}) \ln \frac{\muw^2}{m_{\chi_j}^2} -
\frac49 E_\chi (y_{2j})
\right) \right.\non\\
&+& \frac{\tilde{U}_{j2} \st }{\sqrt{2}\cos\beta} \frac{\mw}{m_{\chi_j}}
\left( t_{2j} G^{\chi,2}_{7} (y_{2j}) + 
t_{2j} \Delta^{\chi,2}_7(y_{2j}) \ln \frac{\muw^2}{m_{\chi_j}^2} \right. \non\\
&-& \left.  \frac43 Y_j R_s F^{(3)}_{7} (y_{2j}) \right) 
 -   \left. \frac89 Y_j  \frac{\mw^2}{m_{\tilde{t}_2}^2} 
t_{2j} \left( R_s + R_b \right) F^{(1)}_7  (y_{2j}) 
   \right]  \label{wc9} \\
\delta^\chi C_{8}^{(1)eff}(\muw)&=& \delta^\chi C_{7}^{(1)eff}(\muw) 
           \left( 7 \rightarrow 8, - \frac49 E_\chi (y_{2j}) \rightarrow
           -\frac16 E_\chi (y_{2j}) \right) . \label{wc10}
\eea
Eq.~(\ref{wc10}) means that  $\delta^\chi C_{8}^{(1)eff}(\muw)$ can be obtained
from the r.h.s.~of eq.~(\ref{wc9}) by replacing  the index 7 with the index 8
in the various functions and by substituting
the term $-4/9 \, E_\chi (y_{2j})$ with $-1/6\,  E_\chi (y_{2j})$.
In eqs.~(\ref{wc9})--(\ref{wc10}) we have used
\bea
G^{\chi,1}_7(x)&=&- \frac89 F^{(1)}_7(x) \left(
\Delta_{t,s}^{(1)}+\Delta_{t,b}^{(1)}
-1 + 3 \ln \frac{m_{\tilde{g}}^2}{m_{\chi_j}^2}\right) \non\\
&+&  {{x\,\left( 85 - 347\,x + 526\,{x^2} \right) }\over 
     {243\,{{\left( 1 - x \right) }^3}}} + 
   {{4\,{x^2}\,\left( -8 + 13\,x + 6\,{x^2} \right) 
       }\over {9\,{{\left( x-1  \right) }^4}}} {\rm Li_2}
\left(1 - {1\over x}\right)
\non\\&-& 
   {{4\,x\left( 20 - 126x + 144\,{x^2} + 39\,{x^3} \right) 
 }\over {81\,{{\left(x -1 \right) }^4}}} \ln x+ 
   {{2\,{x^2}\,\left( 21x-10 \right) }\over 
     {9\,{{\left( x -1\right) }^4}}} \ln^2 x
\label{wc12}
\eea
\bea
G^{\chi,2}_7(x)&=&
- \frac43 F^{(3)}_7(x) \left(\Delta_{t,s}^{(1)}+
\Delta_b^{(1)}+\Delta_{b}^{(2)} -2 
\right)-   {{16\,\left( 3 - 7\,x \right) x}\over 
     {9\,{{\left( x-1 \right) }^3}}}{\rm Li_2}\left(1 - {1\over x}\right) 
\non\\&+& 
  {{4\left(3\,x-5 \right) }\over {9\,{{\left(x-1 \right) }^2}}} 
-   {{4\,\left( 4 - 30\,x + 40\,{x^2} \right)}\over 
     {9\,{{\left( x-1 \right) }^3}}} \ln x- 
   {{16\,\left( 1 - 3\,x \right) x}\over 
     {9\,{{\left( x-1 \right) }^3}}} \ln^2 x 
\label{wc13}
\eea
\bea
G^{\chi,1}_8(x)&=&- \frac89 F^{(1)}_8(x) \left(\Delta_{t,s}^{(1)}+
\Delta_{t,b}^{(1)} - 1 + 3 \ln 
\frac{m_{\tilde{g}}^2}{m_{\chi_j}^2}\right) \non\\
&-& 
  {{x\,\left( 1210 - 437\,x - 1427\,{x^2} \right) }\over 
     {648\,{{\left( x-1 \right) }^3}}} - 
   {{{x^2}\,\left( 49 + 46\,x + 9\,{x^2} \right) 
}\over {12\,{{\left(x-1 \right) }^4}}} {\rm Li_2}\left(1 - {1\over x}\right)
\non\\&-& 
   {{x\,\left( 85 - 603\,x - 387\,{x^2} + 78\,{x^3} \right)}\over 
     {108\,{{\left(x-1 \right) }^4}}}\ln x - 
   {{13\,{x^2}}\over {3\,{{\left( x -1\right) }^4}}}\ln^2 x
\label{wc14}
\eea
\bea
G^{\chi,2}_8(x)&=&
- \frac43 F^{(3)}_8(x) \left(\Delta_{t,s}^{(1)}+
\Delta_b^{(1)}+\Delta_{b}^{(2)} -2 
\right)- 
   {{4\,x\,\left( 3 + 4\,x \right) }\over 
     {3\,{{\left(x-1 \right) }^3}}} {\rm Li_2}\left(1 - {1\over x}\right)  
\non\\&-& 
  {{61 - 39\,x}\over {12\,{{\left(x-1 \right) }^2}}} - 
   {{\left( 7 - 60\,x - 14\,{x^2} \right)}\over 
     {6\,{{\left( x-1 \right) }^3}}}\ln x - 
   {{14\,x}\over {3\,{{\left( x-1 \right) }^3}}}\ln^2 x
\label{wc15}
\eea
\beq
R_i= 3 \ln \frac{\muw^2}{m_{\tilde{g}}^2} + \Delta_{t,i}^{(2)} -1
\label{wc18}
\eeq
\beq 
\Delta^{\chi,1}_7(x) =\frac{32}{27} \left(F^{(1)}_8(x) - 3
F^{(1)}_7(x)  \right) ;~~~~
\Delta^{\chi,2}_7(x) =  \frac{16}{9}  \left(
F^{(3)}_8(x)  -3F^{(3)}_7(x)\right) 
\nonumber
\eeq 
\beq
\Delta^{\chi,1}_8(x) = - \frac{28}{9} F^{(1)}_8(x); ~~~~~~~
\Delta^{\chi,2}_8(x) = - \frac{14}{3}  F^{(3)}_8(x) 
\label{wc17}
\eeq
 The various 
functions $\Delta$ appearing in eqs.~(\ref{wc12})--(\ref{wc18})
contain  the
effect of the renormalization of the chargino-stop-quark vertex 
due to the ${\cal O}(\mug)$ heavy particles. Their explicit expressions are 
given in  appendix A. In the same equations the terms not proportional
to the one-loop functions $F_{7,8}^{(1,3)}$ represent the $\mu$-independent 
part of the contributions coming from two-loop squark-chargino diagrams. 
The  $\mu_\smallw$ dependence in eq.~(\ref{wc9}) and eq.~(\ref{wc10}) 
satisfies the relation
\bea
\frac12 \gamma_0^m \bar{\mt} \frac{\partial C_i^{(0)}}{\partial \bar{\mt} }
&+& \frac{1}{2} \sum_{l=1}^8\gamma_{li}^{(0)eff}C_l^{(0)} = \non \\
&&
\sum_{j=1}^2 \left\{ \frac{\mw^2}{\mst{2}^2} 
\left(  t_{2j}^2   \Delta^{\chi,1}_i (y_{2j})
- \frac{16}3  t_{2j}  Y_j F^{(1)}_i (y_{2j}) 
\right) \right. \non\\
&& + \left.  \frac{\tilde{U}_{j2} \st }{\sqrt{2}\cos\beta} 
\frac{\mw}{m_{\chi_j}}
\left(  t_{2j} \Delta^{\chi,2}_i (y_{2j}) - 4\, Y_j F^{(3)}_{i}(y_{2j}) \right)
\right\}
\label{canc}
\eea
that  ensures that physical observables are independent of
$\muw$ to $O(\al)$. In eq.~(\ref{canc}) $\gamma_0^m = 8$ is the LO anomalous
dimension of the top mass.

Finally, we report the effects of ${\cal O}(\mug)$ heavy particles in diagrams 
involving $W$, physical and unphysical charged scalars exchanges.
As previously discussed,
in our approximation they are introduced as a renormalization
of the relevant  couplings and the corresponding contributions 
 are given by
\bea
\delta^\smallw C_{7}^{(1)eff}(\muw)&=&
\frac43 \left( W_t^b + W_t^s \right) 
G_{7}^{W}\left(tw\right) - 
\frac{23}{27} \left(W_c^b + W_c^s \right)  \\
\delta^\smallw C_{8}^{(1)eff}(\muw)&=&
\frac43 \left( W_t^b + W_t^s \right) 
G_{8}^{W}\left(tw\right) - 
\frac{4}{9} \left(W_c^b + W_c^s \right)  \\
\delta^{\phi_1} C_{7,8}^{(1)eff}(\muw)&=&
\frac{4}{9\, \tan^2 \, \beta} 
\left(H_t^s+H_t^b\right)  F_{7,8}^{(1)}\left(th\right) \non\\
&+& \frac{4}3
\left(H_t^s+H_b\right) F_{7,8}^{(2)} \left(th\right)\\
\delta^{\phi_2} C_{7,8}^{(1)eff}(\muw)&=&
\frac49 \left(U_t^s+U_t^b\right)  F_{7,8}^{(1)}\left(tw\right)
- \frac43 \left(U_t^s+U_b\right) F_{7,8}^{(2)} \left(tw\right) ~~.
\eea
Here $tw = \bar{\mt}(\mu_\smallw)^2/\mw^2, \: 
 th = \bar{\mt}(\mu_\smallw)^2/ \mh^2$,
\bea 
G_{7}^{W}(x)&=&-\frac{23-67x +50x^2}{36(x-1)^3} + \frac{x(2-7x+6x^2)}{
6(x-1)^4} \ln x;\non\\
G_{8}^{W}(x)&=&-\frac{4-5x-5x^2}{12(x-1)^3} +\frac{x(1-2x)}{2(x-1)^4}
\ln x ,
\eea
the functions $F^{(2)}_{7,8}$ are defined in eqs.~(54)--(55) of
ref.~\cite{noi},
and the renormalization functions $W, H, U$ are given in  appendix A.

\section{Numerical Analysis}

In this section we present a numerical analysis of $\br$, with
special attention to the comparison between LO and NLO results in the
supersymmetric model. We focus on the situation in which our approximation
is adequate, {\it i.e.} light stop and charginos, and show how the strong
bounds on the charged Higgs mass can be relaxed. 

We use the same numerical
inputs and conventions as in ref.~\cite{noi}, except for the photon-energy
cut off $E_\gamma > (1-\delta )m_b/2$, which is now chosen to be $\delta=
0.90$ (see discussion in ref.~\cite{kagan}).
In our numerical analysis we also 
include the leading logarithmic electromagnetic effects 
as computed in refs.~\cite{marc,kagan}. 
They amount to a decrease of the calculated
branching ratio of  about $7\%$ to $8\%$. 
We neglect all other two-loop  electroweak effects  which appear
to be very small~\cite{strum2}.

Light supersymmetric particles can affect the 
measured values of the $B^0-\bar B^0$ mixing $\Delta M_B$
and of the CP-violating
$\epsilon_K$
parameter and, ultimately, the extraction of the CKM angles, which enter
the calculation of $\br$.
Within our
assumptions, it is not difficult to compute
this effect. The dominant supersymmetric contributions to 
$\Delta M_B$ and $\epsilon_K$
come from box diagrams
with chargino/light stop and charged Higgs/top quark exchange. 
Under the assumption of
 minimal flavour violation, these contributions have the same structure of
CKM angles as the top-quark box diagram in the SM. Since the dominant
effects on $\Delta M_B$ and $\epsilon_K$ in the SM indeed come
from the top quark, the CKM-angle dependence is the same for all
contributions, and the effect of supersymmetry is parametrized by
\beq
R=1+ \frac{\Delta_{SUSY}}{\Delta_{SM}}~.
\label{rpar}
\eeq
Here $\Delta_{SM}$ is the amplitude of the top-quark box diagram, and
$\Delta_{SUSY}$ is the amplitude of
the chargino/light stop and charged Higgs/top quark box diagrams, whose 
analytical expressions are given in the appendix B. The crucial point is
that $R$ is independent of CKM angles.
In the whole supersymmetric parameter space, $R$ turns out to be
larger than one.

In terms of the Wolfenstein parameters $\eta$ and $\rho$, the CKM structure
of $\Delta M_B$ and $\epsilon_K$ in supersymmetry is
\beq
\Delta M_B \sim |V_{td}^\star V_{tb}|^2~R \sim \left[ (1-\rho )^2 +\eta^2
\right] R
\eeq
\beq
\epsilon_K \sim | {\rm Im} (V_{td}^\star V_{ts})^2| ~R \sim
\eta (1-\rho ) R .
\eeq
The new physics contribution can be effectively reabsorbed in the 
Wolfenstein parameters $(1-\rho )\sqrt{R}$ and $\eta \sqrt{R}$. This means
that the ``true" values of $\eta$ and $\rho$ are related to the values
$\eta_{SM}$ and $\rho_{SM}$ extracted from the usual SM fit by
\beq
\eta = \frac{\eta_{SM}}{\sqrt{R}}
\label{eta}
\eeq
\beq
\rho = 1- \frac{(1-\rho_{SM})}{\sqrt{R}}.
\label{rho}
\eeq
We have checked that eqs.~(\ref{eta})--(\ref{rho}) give an extremely
good approximation of a complete numerical fit.
The combination of the CKM angles entering the evaluation of $\br$ is
\beq
X\equiv \left| \frac{V_{ts}^\star V_{tb}}{V_{cb}}\right|^2=
1+(2\rho -1)\lambda^2 =1+ \left[ 1-\frac{2}{\sqrt{R}}(1-\rho_{SM})\right]
\lambda^2,
\eeq
where $\lambda =|V_{us}|$ does not depend on the supersymmetric masses.
Therefore, in the presence of supersymmetry, the value of $X$ is related to
the usual SM input $X_{SM}$ by
\beq
X=X_{SM} +(\lambda^2+1-X_{SM})\left( 1-\frac{1}{\sqrt{R}}\right) .
\eeq
This is the value of $X$ to be used in a supersymmetric analysis of
$\br$.

It is known that the chargino/stop contribution to the Wilson
coefficient $C_7$ can interfere 
with the SM and charged Higgs contributions
 either constructively or destructively, 
depending on the parameters choice. The case of destructive interference
is particularly interesting, since it is possible to make a light charged
Higgs consistent with
the measurement of $\br$. For illustration, we choose the supersymmetric
parameters corresponding to a strong cancellation between the chargino
and charged-Higgs contributions, and plot in fig.~1 the predicted $\br$
as a function of $\mug$. The LO result in supersymmetry is compared with
the result obtained in the limit $\mug \to \infty$ (denoted by 
{\sf LOdecoupled}
in fig.~1). This shows how the asymptotic behaviour is reached, as $\mug$
grows. The NLO result is also shown in fig.~1. 
In this case, we cannot define an 
asymptotic behaviour, as the NLO branching ratio
hardly decreases with $\mug$ and
 actually starts rising again for $\mug>0.7$ TeV. 
This non-decoupling effect is not surprising~\cite{nond}. 
It appears because, at the
NLO, the chargino vertex contains terms proportional to $\log (\mug / \muw )$.
Indeed, the chargino coupling constant is related to the ordinary gauge and
Yukawa couplings only by supersymmetry. In the effective theory below $\mug$,
supersymmetry is explicitly broken and the renormalization of the chargino
vertex develops a logarithm of the large supersymmetry-breaking scale.
Notice the important effect of the NLO corrections  calculated here.
For the supersymmetric parameters of fig.~1 and for $\mug=1$ TeV, 
the LO branching ratio is 
$1.9 \times 10^{-4}$, while
 the NLO result is $3.7 \times 10^{-4}$. 
Had we neglected in the NLO analysis
our new supersymmetric contributions to the matching conditions, 
the result would have been  
$2.4 \times 10^{-4}$. The large shift induced by the NLO corrections
to $C^{eff}_{7,8}(\muw)$ is mainly due by
 the large non-decoupling logarithms
of $\mug / \muw $.
Finally, in fig.~1 we also show the LO and NLO results obtained in a 
two-Higgs doublet model with charged-Higgs mass equal to the supersymmetric
case. The light charged Higgs (taken with a mass of 100 GeV in fig.~1) is
clearly incompatible with the measurement of $\br$, in the absence of an
appropriate chargino contribution. 

In order to assess the impact of the different contributions, 
it is also interesting to
consider the NLO Wilson coefficients $C^{eff}_{7,8}$ at the weak scale
$\muw =M_W$ in the supersymmetric configuration chosen in fig.~1 
with all heavy masses set to
1 TeV, and compare them with the corresponding quantities in the SM and the
two-Higgs doublet model. The results are shown in the table 1. Notice that,
in the supersymmetric case, the QCD corrections to $C^{eff}_{7}$ 
amount to about 60{\%}. 
Such a large correction is clearly related to the fact that the LO approximate 
cancellation among different contributions is partially spoiled at the NLO
level. Indeed, in the case under consideration, the NLO effects
increase the SM contribution to $|C_7^{eff} (M_W)|$ by almost 10\% and 
decrease the charged-Higgs and the chargino contributions by about 
20\% and 30\%, respectively.
This situation of partial cancellation and enhanced sensitivity to NLO
correction is actually the most interesting phenomenologically. It is
under this condition that a light charged-Higgs mass is still allowed
and that our calculation is essential.

%%%%%%%%%%%%%%%%%%% table 1
\renewcommand{\arraystretch}{1.1}
\begin{table}[h]
\begin{center}
\begin{tabular}{|c||c|c|} 
\hline 
  &  $C^{eff}_{7}(M_W)$ & $ C^{eff}_{8}(M_W)$   \\ \hline \hline 
LO SM         &-0.198 &   -0.098    \\ \hline 
NLO  SM          & -0.220 & -0.119      \\ \hline
LO 2HDM         & -0.529  &  -0.336    \\ \hline 
NLO 2HDM         & -0.493&    -0.326   \\ \hline 
LO SUSY         &-0.143 &    -0.141  \\ \hline 
NLO SUSY         & -0.229  &   -0.183    \\ \hline 
\end{tabular} 
\caption{\sf Wilson coefficients evaluated at the scale $\mw$ in 
the Standard Model (SM), the two-Higgs doublet model (2HDM) and 
supersymmetry (SUSY)
at LO and at NLO for
$\tan\beta=1$, $\mh=\mst{2}=m_{{\chi}_2}=
100$ GeV, $m_{{\chi}_1}=300$ GeV,
$\theta_{\tilde{t}}=-\pi/10$, $A_b=A_t$;
  all other squarks and the gluino are
degenerate with mass of  1 TeV. The lighter chargino is predominantly 
higgsino.}
\end{center} 
\end{table} 

We now want to investigate the region of parameter space in which we can
significantly relax
the bound
on the charged Higgs mass from $\br$. We have scanned over the relevant
supersymmetric parameters, assuming a stop mixing angle less than
$\pi/10$ in absolute value, consistently with our assumption of a mainly
right-handed light stop. In fig.~2
we show,  for $\tan \beta=2 $ and 4,
the maximum values of the lighter chargino
and lighter stop masses for which a charged Higgs of 100 GeV
is consistent with the 
95\% C.L.~CLEO limit of $4.2\times 10^{-4}$. We have chosen this value
to give a conservative estimate. With improved experimental results and
revised analyses of the photon energy spectrum, this bound may become
significantly more restrictive.
The effect of the NLO corrections is to make
the upper bounds on the chargino and stop masses quite more stringent.
In the same figure we also present
the results obtained  using the  renormalization group evolution
and the SM and charged-Higgs contributions to the Wilson coefficients 
at the NLO, 
while the purely supersymmetric effects on
$C^{eff}_{7,8}(\muw )$ are evaluated at the LO. This is to 
show the impact of the
calculation presented here. As seen from the figure, the effect of the
QCD corrections to the supersymmetric contribution to $C^{eff}_{7,8}(\muw )$
is very significant.

\section{Conclusions}

In this paper we have computed the QCD NLO corrections to the matching
conditions of the Wilson coefficients relevant for $\br$. 
We have used some theoretical assumptions to simplify the result and to
concentrate on the most relevant part of the supersymmetric parameter
space. We have assumed minimal flavour violation or, in other words, that
the flavour-violating interactions in supersymmetry are primarily in the
charged-current sector and are determined by the CKM angles. This is often
a good approximation in a variety of models. At worst, our result represents
an unavoidable contribution to be added to new effects coming from other
sources of flavour violation. This assumption allow us also to limit the
number of unknown parameters.

We have focused on the case in which the purely supersymmetric contribution
is sizeable and can compensate the effect of a light charged-Higgs boson.
For this reason, we have assumed that charginos and a mainly right-handed
stop are considerably lighter than the other squarks and the gluino. This
assumption has allowed us to use an effective theory, in which the heavy
particles are integrated out, retaining only the first term in an
expansion in the ratio between light and heavy masses.

In sect.~3 we have presented the analytic results for the NLO matching 
conditions of the Wilson coefficients in supersymmetry, under the 
approximations stated above. We have also performed a numerical investigation
of the supersymmetric parameter region 
in which there is a sizeable cancellation among various different terms. 
In this region the impact of the NLO corrections to the supersymmetric
contribution
to the Wilson
coefficients is very significant because of  two conspiring effects.
On one side there is a large renormalization
of the one-loop supersymmetric
contribution, mostly coming from logarithms of the ratio
between the high and the intermediate mass scales. On the other side, 
the effect of the large NLO corrections to the supersymmetric contribution
is greatly enhanced whenever there is an approximate cancellation at the LO.

\vspace{1cm}
We would like to thank  A.~Buras, G.~Martinelli,
M.~Misiak, M.~Neubert, and 
L.~Silvestrini for useful discussions.
Right before submitting this paper, we learned from 
M.~Misiak and J. Urban that they have reproduced our results in
the limit of infinite gluino mass. 

\appendix
\setcounter{section}{1}
\begin{appendletter}

\section*{Appendix A}
In this appendix we present the results  for
the renormalized vertices in the effective theory at scales between
$\mug$ and $\muw$. In deriving the various corrections we exploited the 
fact 
that, in our framework, the light stop is mainly right-handed and 
therefore we can
neglect terms  $O(\sin \theta_{\tilde{t}}\,  \muw / \mug)$. 
However in the formulae below we do not expand functions of $\mst{2}^2 / \mg^2$
and we keep the explicit $\sin^2 \theta_{\tilde{t}}$ terms, because we feel
it will be easier for the reader to understand the origin of the different
terms.
\subsection*{W}
The renormalized $\bar{d}\, u \, W$ vertex is
$$ (-i) \frac{g}{\sqrt{2}}\, V_{ud} \,\gamma^\mu \, \am 
\left( 1 + \, \frac{\al}{3 \pi} \,W_u^d \right) $$
where
$$ W_u^d = \frac12 \left( \cos^2 \theta_{\tilde{u}}\,
          W[x_1,w_1] + \sin^2 \theta_{\tilde{u}} \,
   W[x_1,w_2] \right) $$
with $x_1 = \msd{1}^2 /\mg^2 $, $w_i = \msu{i}^2 / \mg^2 $ and
\bea
W[x,y] & = &  
\frac{x + y - 2 xy}{(x-1)(y-1)} +
\frac{x^3 -2 x y +x^2 y}{(x-1)^2(x-y)} \ln\, x
+ \frac{2 x y -x y^2 - y^3}{(y-1)^2 (x-y)} \ln\, y . \nonumber
\eea
For equal masses $W[x,x] =0$.

\subsection*{Unphysical scalar}
The renormalized   $\bar{d}\, t \, \phi^+$ vertex is
$$ (-i) \frac{g}{\sqrt{2}\, \mw}\, V_{td}\, 
\left[ m_d \,\am \left( 1 + \frac{\al}{3 \pi} \, U_d \right)
- \mt\, \ap \left( 1 + \frac{\al}{3 \pi} \,U_t^d \right)  \right] $$
where
\bea
U_d & = &  \frac12 \left( H_1[x_1] - \cdt\, H_1[u_1]
         -\sdt \,H_1[u_2] \right) \nonumber \\ &&
 -  2\, \frac{ A_d - \mu\, \tan \beta}{\mg} 
  H_2[x_1,x_2] \nonumber \\ &&
+2\, \frac{ A_d - \mu\, \tan \beta }{\mg}
\left( \cdt \,H_2[u_1,x_2] + \sdt\,  H_2[u_2,x_2]
\right), \nonumber \\ 
U_t^d & = &  - \frac12 \left( H_1[x_1] - \cdt\, H_1[u_1]
         -\sdt \,H_1[u_2] \right) \nonumber \\ &&
-  2 \, \frac{ A_u - \mu\, \cot \beta }{\mg} H_2[u_1,u_2]
 \nonumber \\ &&
+2\, \frac{ A_u - \mu\, \cot \beta }{\mg}
\left( \cdt \,H_2[u_2,x_1] + \sdt\,  H_2[u_1,x_1]
\right). \nonumber
\eea
with $x_2 = \msd{2}^2 /\mg^2 $, $u_i = \mst{i}^2 / \mg^2 $ and
\bea
H_1[x] & = & \frac1{1-x} + \frac{2\, x - x^2}{(1-x)^2} \ln\, x
\nonumber
\eea
\bea
H_2[x,y] & = &  
\frac{x}{(1-x)(x-y)} \ln\, x + \frac{y}{(1-y) (y-x)} \ln\, y
\nonumber
\eea
\subsection*{Physical scalar}
We write the renormalized vertex $\bar{d}\, t \, h^+$ as
$$  \frac{i\,g}{\sqrt{2}\, \mw}\, V_{td}\, 
   \left[\tan  \beta \, m_d \,\am \left( 1 + \frac{\al}{3 \pi} 
   \, H_d \right) + \cot \beta \, \mt\, \ap 
\left( 1 + \frac{\al}{3 \pi} \,H_t^d \right) \right] $$
where
\bea
H_d & = &  \frac12 \left( H_1[x_1] - \cdt\, H_1[u_1]
         -\sdt \,H_1[u_2] \right)  \nonumber \\ &&
 -  2\, \frac{ A_d - \mu\, \tan \beta }{\mg} 
  H_2[x_1,x_2] \nonumber \\ &&
+2\, \frac{ A_d + \mu\, \cot \beta }{\mg}
\left( \cdt \,H_2[u_1,x_2] + \sdt\,  H_2[u_2,x_2]
\right), \nonumber \\ 
H_t^d & = &  - \frac12 \left( H_1[x_1] - \cdt\, H_1[u_1]
         -\sdt \,H_1[u_2] \right)  \nonumber \\ &&
-  2 \, \frac{ A_u - \mu\, \cot \beta }{\mg} H_2[u_1,u_2]
 \nonumber \\ &&
+2\, \frac{ A_u + \mu\, \tan \beta }{\mg}
\left( \cdt \,H_2[u_2,x_1] + \sdt\,  H_2[u_1,x_1]
\right). \nonumber
\eea
\subsection*{Chargino}
Unlike the case of the $W$ and of physical and unphysical scalars, in the 
chargino sector the $O(\al)$ corrections due to gluon and gluino exchanges
are not separately finite. Therefore, as shown below, the renormalization
of the chargino vertex due to the $O(\mug)$ heavy particles will contain
$1/\epsilon$ poles ($\epsilon = (4 - n)/2$, with $n$ the dimension of the 
space-time). The cancellation of these poles against similar terms coming
from two-loop diagrams in which a gluon is present provides  a check of our
calculation. We write the corrected $\bar{\chi}_a \, d \, \tilde{t}_2$ 
vertex as
\bea &(-i g)& \! \! \! V_{td} \left\{ -\st \, \widetilde{U}_{a2} \Yd \ap 
\left( 1 + \frac{\al}{3 \pi} \,  C_d \right)  \right.   \non \\
 &+&  \!\hspace{-.7cm} \left .  
  \left[ \ct \, \widetilde{V}_{a2} \Yt a_-
\left( 1 + \frac{\al}{3 \pi} \, 
                C^a_{t,d} \right)
+ \st\, \widetilde{V}_{a1} \am  \left( 1 + \frac{\al}{3 \pi} \, 
   C^b_{t,d} \right)  \right]  \right\} 
\eea
with ($\bar{\mu}$ is the 't-Hooft mass)
\bea
C_d &=& 
 \frac{3}{2\, \epsilon}
  - \frac32 \ln \left( \frac{\mg^2}{\bar{\mu}^2} \right) + \Delta_d^{(1)} +
     \Delta_d^{(2)} + \eta_Y \nonumber \\ 
C^a_{t,d} &=&
 \frac3{2\, \epsilon}
  - \frac32 \ln \left( \frac{\mg^2}{\bar{\mu}^2} \right)+\Delta_{t,d}^{(1)} + 
   \Delta_{t,d}^{(2)} + \eta_Y     \nonumber \\ 
C^b_{t,d} &=&
 - \frac3{2\, \epsilon} + \frac32 \, \ln \left( \frac{\mg^2}{\bar{\mu}^2} 
 \right) + \Delta_{t,d}^{(1)} + \eta_g .
\label{vertcharg}
\eea
The  $\Delta^{(i)}_{t(,d)}$ functions in eq.~(\ref{vertcharg}), also 
appearing in eqs.~(\ref{wc12})--(\ref{wc18}), are given by 
\bea
\Delta_d^{(1)} & = & - \frac34 - \frac12 H_1[x_2] 
\nonumber \\
\Delta_{t,d}^{(1)} & = & - \frac34 - \frac12 H_1[x_1] 
\nonumber \\
\Delta_d^{(2)} & = & \frac52 +  \left( 1 -
\cot \theta_{\tilde{t}}\,  \frac{\mt\,\mg}{\msd{2}^2} \right) H_3[x_2] 
+  \frac12 H_1[x_1] + \frac12 H_1[x_2] \nonumber \\ 
&& -  2 \, \frac{A_d - \mu\, \tan \beta}{\mg} H_2[x_1,x_2] 
 \nonumber \\
\Delta_{t,d}^{(2)} & = & \frac52 + H_3[x_1]
 + \frac12 H_1[u_1] + \frac12 H_1[u_2] 
-  2 \, \frac{A_u - \mu\, \cot \beta}{\mg} H_2[u_1,u_2] 
\nonumber
\eea
where
\bea
H_3[x] & = & \frac{2 x}{1-x} \ln\, x . \nonumber
\eea
In eq.~(\ref{vertcharg}) the factors $\eta_Y = -3/2$ and 
$\eta_g = -1/2$ are induced by
the fact that the $\MS$ renormalization does not preserve supersymmetry,
as discussed in the text.
\end{appendletter}

\appendix
\setcounter{section}{2}
\begin{appendletter}

\section*{Appendix B}
This appendix contains the expression of the parameter $R$ 
defined in eq.~(\ref{rpar}), which can be derived from the results
of ref.~\cite{bert}. The SM contribution is
\bea
\Delta_{SM}= \frac{tw^3 - 12 \,tw^2 + 15 \,tw -4 + 6 \,tw^2\, \ln tw}{4\,(tw-1)^3}
\eea
with $tw = \bar{\mt}(\mu_\smallw)^2/\mw^2$.
The supersymmetric contribution can be split into charged Higgs and chargino 
contributions:
\bea
\Delta_{SUSY}= \Delta_H+\tilde{\Delta}.
\eea
\bea
\Delta_H=\cot^4 \beta \frac{(th^2 -1 -2 \,th\, \ln th)\,th}{4 \,(th-1)^3}
+ 2 \cot^2 \beta \,tw \left[ F'(tw,hw) + \frac1{4} G'(tw,hw)\right]
\eea
with $ th = \bar{\mt}(\mu_\smallw)^2/ \mh^2$ and $hw=\mh^2/\mw^2$. The
functions $F'$ and $G'$ are given by
\bea
F'(x,y)&=& \frac{(x^2-y)\ln x}{(x-y)^2(x-1)^2}-
\frac{y\ln y}{(x-y)^2(y-1)}-\frac{1}{(x-y)(x-1)}
 \non\\
G'(x,y)&=&  
\frac{x}{(x-y)(x-1)}\left[ 1-\left( \frac{1}{x-1}+\frac{y}{x-y}\right) \ln x
\right] +\frac{y^2\ln y}{(x-y)^2(y-1)}
\eea
Consistently with our assumptions,
we keep only the light stop 
contributions to $\tilde{\Delta}$:
\bea 
\tilde{\Delta}= \sum_{i,j=1,2} \frac{\mw^4}{\mt^2 m_{\chi_j}^2}
\,t_{2i}^2 \,t_{2j}^2  \, G'(y_{2j}, x_{ij}),
\eea
where  $y_{2j}$ is the variable defined in eq.~(\ref{wc3}) and  
$x_{ij} $ is the ratio of the squared masses
 of two charginos.

\end{appendletter}

\newpage
\begin{figure}[t]
\vspace{-2cm}
\centerline{
\psfig{figure=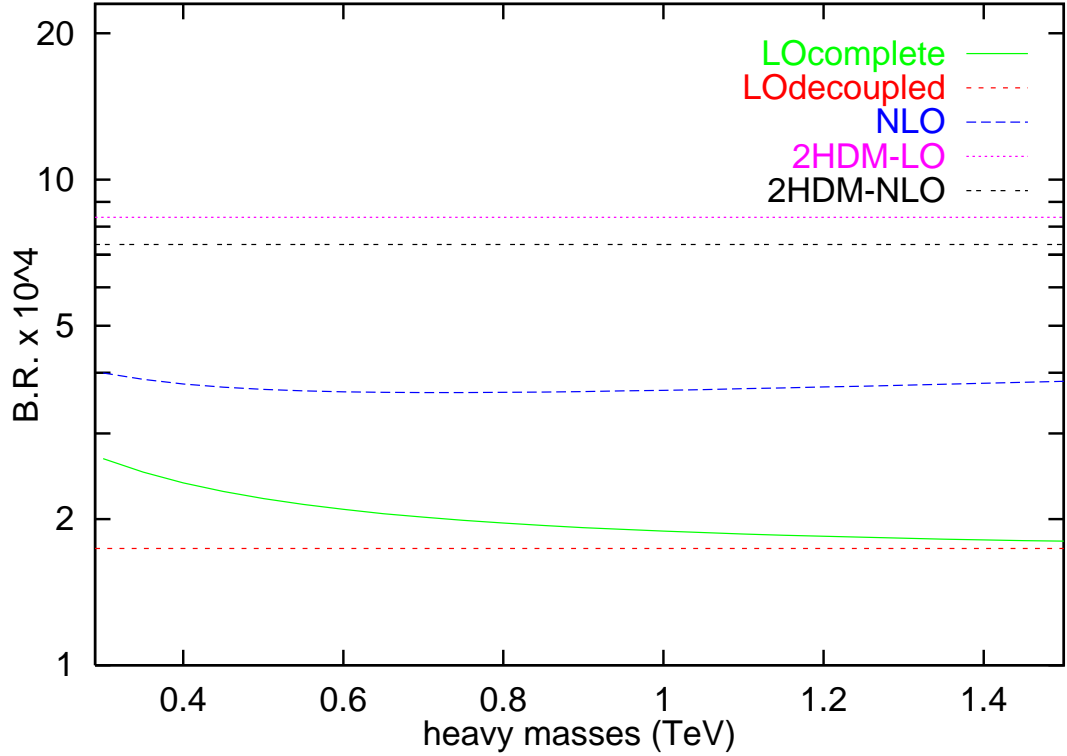,height=4 in,rheight=4.1 in}
  }
\caption{\sf Dependence of the prediction for the $B\to X_s \gamma$
 branching ratio 
on the heavy mass scale $\mug$ for the following choice of parameters:
$\tan\beta=1$, $\mh=\mst{2}=m_{{\chi}_2}=
100$ GeV, $m_{{\chi}_1}=300$ GeV, 
$\theta_{\tilde{t}}=-\pi/10$, $A_b=A_t$, all heavy particle masses
equal to $\mug$; the lighter chargino is predominantly 
higgsino.  
The value corresponding to "LOdecoupled" is the result
of the LO calculation in the limit in which the heavy masses decouple.
The two upper lines are the LO and NLO predictions in the two-Higgs 
doublet model (2HDM).
}
\label{fig1}
\end{figure}
\newpage
\begin{figure}[t]
\vspace{-2cm}
\centerline{
\psfig{figure=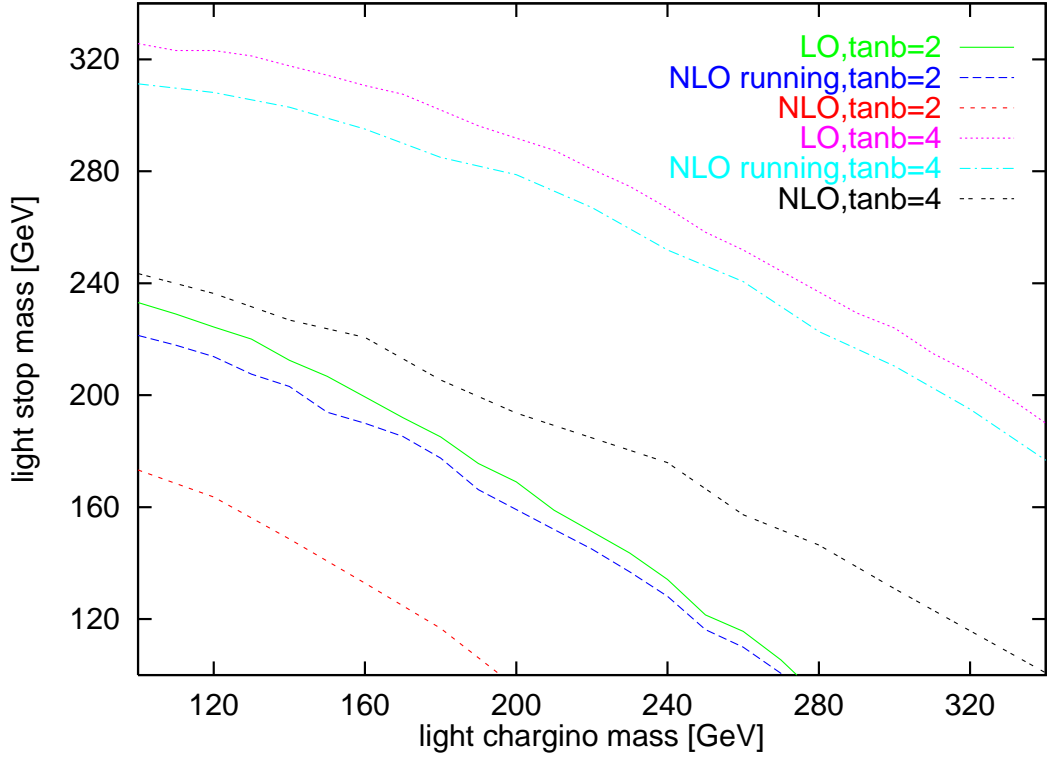,height=4 in,rheight=4.1 in}
  }
\caption{\sf Upper bounds on the lighter chargino and stop masses 
from the CLEO 95\% CL limit on $\br$ in the case $\mh=100$ GeV. 
We have taken
$|\theta_{\tilde t}|<\pi/10$,  $|\mu|<500$ GeV, $A_b=A_t$, and set
all heavy masses to 1 TeV. 
For  $\tan\beta=2$ and $4$ we show
the results of the LO  and
 NLO calculations. The result of 
neglecting the new NLO supersymmetric contributions to the 
Wilson coefficients is 
labelled as "NLO running".
}
\label{fig2}
\end{figure}

\end{document}